\documentclass{svmult}

\usepackage{makeidx}     
\usepackage{graphicx}    
\usepackage{multicol}    
\usepackage{amssymb}
\usepackage{float}
\usepackage{amsmath}

\makeindex             

\begin{document}

\title*{Single-shot read-out of an individual electron spin in a quantum dot}
\author{J.M. Elzerman\and
R. Hanson\and
L.H. Willems van Beveren\and
B. Witkamp\and
L.M.K. Vandersypen\and
L.P. Kouwenhoven
}
\authorrunning{J.M. Elzerman \emph{et al.}}
\institute{\emph{Kavli Institute of Nanoscience Delft and ERATO Mesoscopic Correlation Project,\\
Delft University of Technology, PO Box 5046, 2600 GA Delft, The Netherlands}}

\maketitle

\def\ua{\uparrow}
\def\da{\downarrow}

\bigskip
\bigskip

\noindent Spin is a fundamental property of all elementary particles. Classically it can be viewed as a tiny magnetic moment, but a measurement of an electron spin along the direction of an external magnetic field can have only two outcomes: parallel or anti-parallel to the field \cite{sakurai}. This discreteness reflects the quantum mechanical nature of spin. Ensembles of many spins have found diverse applications ranging from magnetic resonance imaging \cite{wehrli} to magneto-electronic devices \cite{wolf}, while individual spins are considered as carriers for quantum information. Read-out of single spin states has been achieved using optical techniques \cite{blatt}, and is within reach of magnetic resonance force microscopy \cite{mamin}. However, electrical read-out of single spins [6-13] has so far remained elusive. Here, we demonstrate electrical single-shot measurement of the state of an individual electron spin in a semiconductor quantum dot \cite{lpk01}. We use spin-to-charge conversion of a single electron confined in the dot, and detect the single-electron charge using a quantum point contact; the spin measurement visibility is $\sim 65\%$. Furthermore, we observe very long single-spin energy relaxation times (up to $\sim 0.85$ ms at a magnetic field of 8 Tesla), which are encouraging for the use of electron spins as carriers of quantum information.

\newpage
\def\ua{\uparrow}
\def\da{\downarrow}

\section{Measuring electron spin in quantum dots}
\label{spin}
In quantum dot devices, single electron charges are easily measured. Spin states in quantum dots, however, have only been studied by measuring the average signal from a large ensemble of electron spins [17-22]. In contrast, the experiment presented here aims at a single-shot measurement of the spin orientation (parallel or antiparallel to the field, denoted as spin-$\ua$ and spin-$\da$, respectively) of a particular electron; only one copy of the electron is available, so no averaging is possible. The spin measurement relies on spin-to-charge conversion \cite{fujisawa02,rh03} followed by charge measurement in a single-shot mode \cite{lu,fujisawa04}. Figure~\ref{fig5:principle}a schematically shows a single electron spin confined in a quantum dot (circle). A magnetic field is applied to split the spin-$\ua$ and spin-$\da$ states by the Zeeman energy. The dot potential is then tuned such that if the electron has spin-$\da$ it will leave, whereas it will stay on the dot if it has spin-$\ua$. The spin state has now been correlated with the charge state, and measurement of the charge on the dot will reveal the original spin state.

\section{Implementation}
\label{implementation}
\begin{figure}[t]
\centering
\includegraphics[width=11.5cm, clip=true]{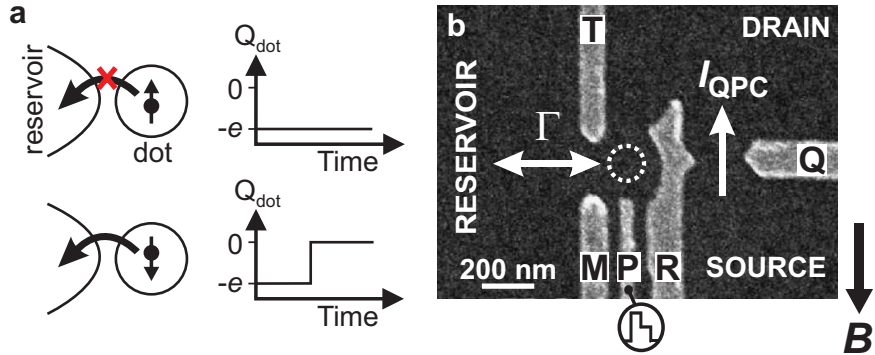}
\caption{Spin-to-charge conversion in a quantum dot coupled to a quantum point contact.
\textbf{(a)} Principle of spin-to-charge conversion. The charge on the quantum dot, $Q_{dot}$, remains constant if the electron spin is $\ua$, whereas a spin-$\da$ electron can escape, thereby changing $Q_{dot}$. 
\textbf{(b)} Scanning electron micrograph of the metallic gates on the surface of a GaAs/Al$_{0.27}$Ga$_{0.73}$As heterostructure containing a two-dimensional electron gas (2DEG) 90 nm below the surface. The electron density is ${2.9\times 10^{15}}$ m${^{-2}}$. (Only the gates used in the present experiment are shown, the complete device is described in Ref.~\protect\cite{jme03}.) Electrical contact is made to the QPC source and drain and to the reservoir via Ohmic contacts. With a source-drain bias voltage of 1 mV, $I_{QPC}$ is about 30 nA, and an individual electron tunnelling on or off the dot changes $I_{QPC}$ by $\sim0.3$ nA. The QPC-current is sent to a room temperature current-to-voltage convertor, followed by a gain 1 isolation amplifier, an AC-coupled 40 kHz SRS650 low-pass filter, and is digitized at a rate of $2.2\times 10^6 $ samples/s. With this arrangement, the step in $I_{QPC}$ resulting from an electron tunnelling is clearly larger than the rms noise level, provided it lasts at least 8 $\mu$s. A magnetic field, ${B}$, is applied in the plane of the 2DEG.}
\label{fig5:principle}
\end{figure}

This concept is implemented using a structure \cite{jme03} (Fig.~\ref{fig5:principle}b) consisting of a quantum dot in close proximity to a quantum point contact (QPC). The quantum dot is used as a box to trap a single electron, and the QPC is operated as a charge detector in order to determine whether the dot contains an electron or not. The quantum dot is formed in the two-dimensional electron gas (2DEG) of a GaAs/AlGaAs heterostructure by applying negative voltages to the metal surface gates $M$, $R$, and $T$. This depletes the 2DEG below the gates and creates a potential minimum in the centre, that is, the dot (indicated by a dotted white circle). We tune the gate voltages such that the dot contains either zero or one electron (which we can control by the voltage applied to gate $P$). Furthermore, we make the tunnel barrier between gates $R$ and $T$ sufficiently opaque that the dot is completely isolated from the drain contact on the right. The barrier to the reservoir on the left is set \cite{jme04} to a tunnel rate $\Gamma \approx (0.05$ ms$)^{-1}$. When an electron tunnels on or off the dot, it changes the electrostatic potential in its vicinity, including the region of the nearby QPC (defined by $R$ and $Q$). The QPC is set in the tunnelling regime, so that the current, $I_{QPC}$, is very sensitive to electrostatic changes \cite{field}. Recording changes in $I_{QPC}$ thus permits us to measure on a timescale of about 8 $\mu$s whether an electron resides on the dot or not \cite{LMKV}. In this way the QPC is used as a charge detector with a resolution much better than a single electron charge and a measurement timescale almost ten times shorter than $1/\Gamma$.

The device is placed inside a dilution refrigerator, and is subject to a magnetic field of 10 T (unless noted otherwise) in the plane of the 2DEG. The measured Zeeman splitting in the dot \cite{rh03}, $\Delta E_Z \approx 200 \mu$eV, is larger than the thermal energy (25 $\mu$eV) but smaller than the orbital energy level spacing (1.1 meV) and the charging energy (2.5 meV).

\section{Two-level pulse technique}
\label{2levelpulse}
\begin{figure}[htbp]
\centering
\includegraphics[width=10.1cm, clip=true]{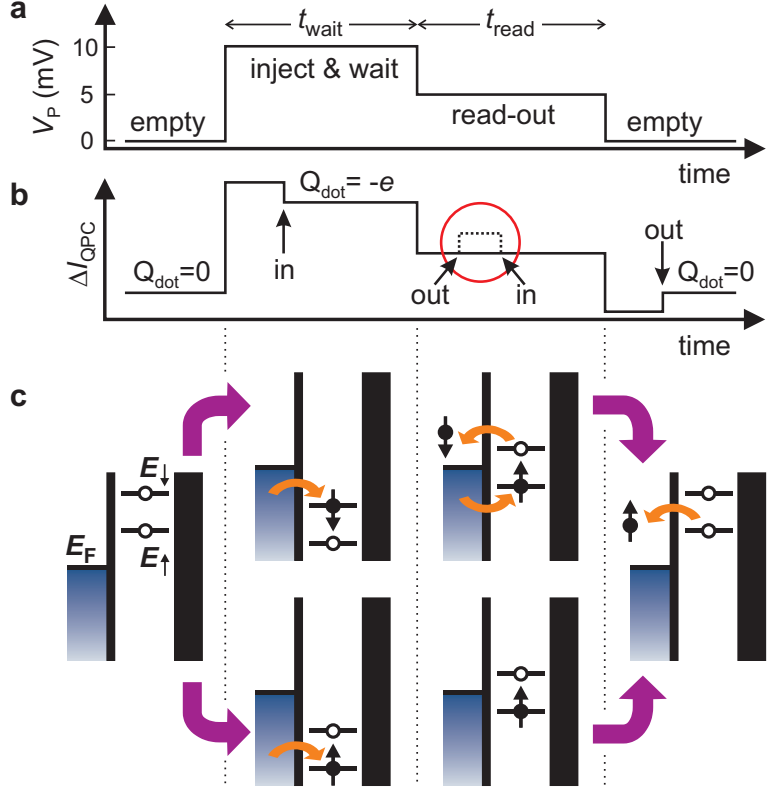}
\caption{Two-level pulse technique used to inject a single electron and measure its spin orientation. 
\textbf{(a)} Shape of the voltage pulse applied to gate $P$. The pulse level is 10 mV during $t_{wait}$ and 5 mV during $t_{read}$ (which is 0.5 ms for all measurements). 
\textbf{(b)} Schematic QPC pulse-response if the injected electron has spin-$\ua$ (solid line) or spin-$\da$ (dotted line; the difference with the solid line is only seen during the read-out stage). Arrows indicate the moment an electron tunnels into or out of the quantum dot. 
\textbf{(c)} Schematic energy diagrams for spin-$\ua$ ($E_{\ua}$) and spin-$\da$ ($E_{\da}$) during the different stages of the pulse. Black vertical lines indicate the tunnel barriers. The tunnel rate between the dot and the QPC-drain on the right is set to zero. The rate between the dot and the reservoir on the left is tuned to a specific value, $\Gamma$. If the spin is $\ua$ at the start of the read-out stage, no change in the charge on the dot occurs during $t_{read}$. In contrast, if the spin is $\da$, the electron can escape and be replaced by a spin-$\ua$ electron. This charge transition is detected in the QPC-current (dotted line inside red circle in (b)).}
\label{fig5:diagrams}
\end{figure}

To test our single-spin measurement technique, we use an experimental procedure based on three stages: (1) empty the dot, (2) inject one electron with unknown spin, and (3) measure its spin state. The different stages are controlled by voltage pulses on gate $P$ (Fig.~\ref{fig5:diagrams}a), which shift the dot's energy levels (Fig.~\ref{fig5:diagrams}c). Before the pulse the dot is empty, as both the spin-$\ua$ and spin-$\da$ levels are above the Fermi energy of the reservoir, $E_F$. Then a voltage pulse pulls both levels below $E_F$. It is now energetically allowed for an electron to tunnel onto the dot, which will happen after a typical time $\sim \Gamma^{-1}$. The particular electron can have spin-$\ua$ (shown in the lower diagram) or spin-$\da$ (upper diagram). (The tunnel rate for spin-$\ua$ electrons is expected to be larger than that for spin-$\da$ electrons \cite{rh04}, i.e. $\Gamma_{\ua} > \Gamma_{\da}$, but we do not assume this a priori.) During this stage of the pulse, lasting $t_{wait}$, the electron is trapped on the dot and Coulomb blockade prevents a second electron to be added. After $t_{wait}$ the pulse is reduced, in order to position the energy levels in the read-out configuration. If the electron spin is $\ua$, its energy level is below $E_F$, so the electron remains on the dot. If the spin is $\da$, its energy level is above $E_F$, so the electron tunnels to the reservoir after a typical time $\sim \Gamma_{\da}^{-1}$. Now Coulomb blockade is lifted and an electron with spin-$\ua$ can tunnel onto the dot. This occurs on a timescale $\sim \Gamma_{\ua}^{-1}$ (with $\Gamma=\Gamma_{\ua}+\Gamma_{\da}$). After $t_{read}$, the pulse ends and the dot is emptied again.

The expected QPC-response, $\Delta I_{QPC}$, to such a two-level pulse is the sum of two contributions (Fig.~\ref{fig5:diagrams}b). First, due to a capacitive coupling between pulse-gate and QPC, $\Delta I_{QPC}$ will change proportionally to the pulse amplitude. Thus, $\Delta I_{QPC}$ versus time resembles a two-level pulse. Second, $\Delta I_{QPC}$ tracks the charge on the dot, i.e. it goes up whenever an electron tunnels off the dot, and it goes down by the same amount when an electron tunnels on the dot. Therefore, if the dot contains a spin-$\da$ electron at the start of the read-out stage, $\Delta I_{QPC}$ should go up and then down again. We thus expect a characteristic step in $\Delta I_{QPC}$ during $t_{read}$ for spin-$\da$  (dotted trace inside red circle). In contrast, $\Delta I_{QPC}$ should be flat during $t_{read}$ for a spin-$\ua$ electron. Measuring whether a step is present or absent during the read-out stage constitutes our spin measurement.

\begin{figure}[htbp]
\centering
\includegraphics[width=11.5cm, clip=true]{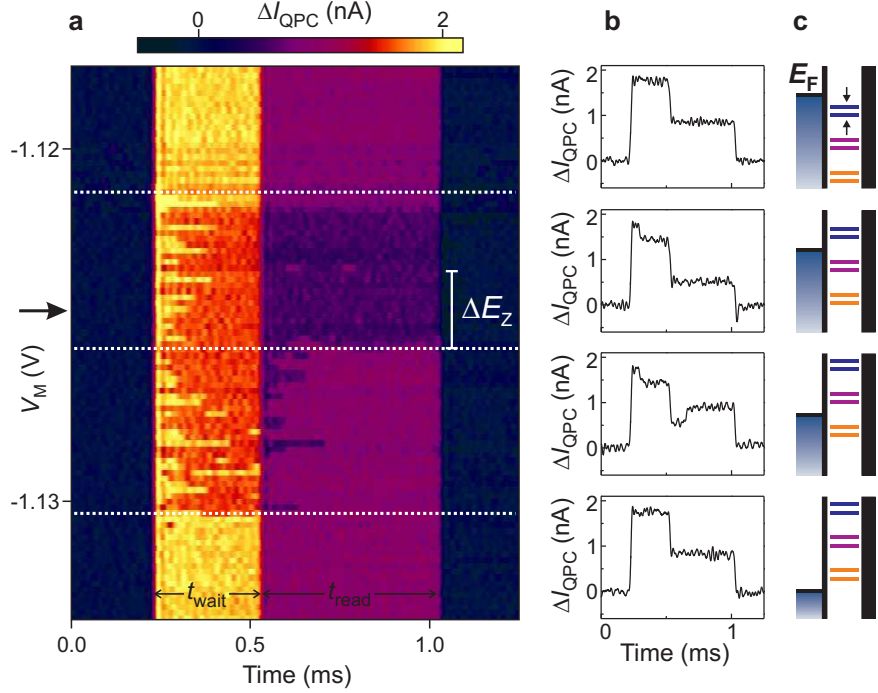}
\caption{Tuning the quantum dot into the spin read-out configuration. We apply a two-stage voltage pulse as in Fig.~\ref{fig5:diagrams}a ($t_{wait}$ = 0.3 ms, $t_{read}$ = 0.5 ms), and measure the QPC-response for increasingly negative values of $V_M$. 
\textbf{(a)} QPC-response (in colour-scale) versus $V_M$. Four different regions in $V_M$ can be identified (separated by white dotted lines), with qualitatively different QPC-responses. 
\textbf{(b)} Typical QPC-response in each of the four regions. This behaviour can be understood from the energy levels during all stages of the pulse. 
\textbf{(c)} Schematic energy diagrams showing $E_{\ua}$ and $E_{\da}$ with respect to $E_F$ before and after the pulse (blue), during $t_{wait}$ (orange) and during $t_{read}$ (purple), for four values of $V_M$. For the actual spin read-out experiment, $V_M$ is set to the optimum position (indicated by the arrow in a).}
\label{fig5:tuning}
\end{figure}

\section{Tuning the quantum dot into the read-out configuration}
\label{tune}
To perform spin read-out, $V_M$ has to be fine-tuned so that the position of the energy levels with respect to $E_F$ is as shown in Fig.~\ref{fig5:diagrams}c. To find the correct settings, we apply a two-level voltage pulse and measure the QPC-response for increasingly negative values of $V_M$ (Fig.~\ref{fig5:tuning}a). Four different regions in $V_M$ can be identified (separated by white dotted lines), with qualitatively different QPC-responses. The shape of the typical QPC-response in each of the four regions (Fig.~\ref{fig5:tuning}b) allows us to infer the position of $E_{\ua}$ and $E_{\da}$ with respect to $E_F$ during all stages of the pulse (Fig.~\ref{fig5:tuning}c). 

In the top region, the QPC-response just mimics the applied two-level pulse, indicating that here the charge on the dot remains constant throughout the pulse. This implies that $E_{\ua}$ remains below $E_F$ for all stages of the pulse, thus the dot remains occupied with one electron. In the second region from the top, tunnelling occurs, as seen from the extra steps in $\Delta I_{QPC}$. The dot is empty before the pulse, then an electron is injected during $t_{wait}$, which escapes after the pulse. This corresponds to an energy level diagram similar to before, but with $E_{\ua}$ and $E_{\da}$ shifted up due to the more negative value of $V_M$ in this region. In the third region from the top, an electron again tunnels on the dot during $t_{wait}$, but now it can escape already during $t_{read}$, irrespective of its spin. Finally, in the bottom region no electron-tunneling is seen, implying that the dot remains empty throughout the pulse. 

Since we know the shift in $V_M$ corresponding to shifting the energy levels by $\Delta E_Z$ \cite{jme04}, we can set $V_M$ to the optimum position for the spin read-out experiment (indicated by the arrow). For this setting, the energy levels are as shown in Fig.~\ref{fig5:diagrams}c, i.e. $E_F$ is approximately in the middle between $E_{\ua}$ and $E_{\da}$ during the read-out stage.

\section{Single-shot read-out of one electron spin}
\label{exp}
\begin{figure}[htbp]
\centering
\includegraphics[width=11.5cm, clip=true]{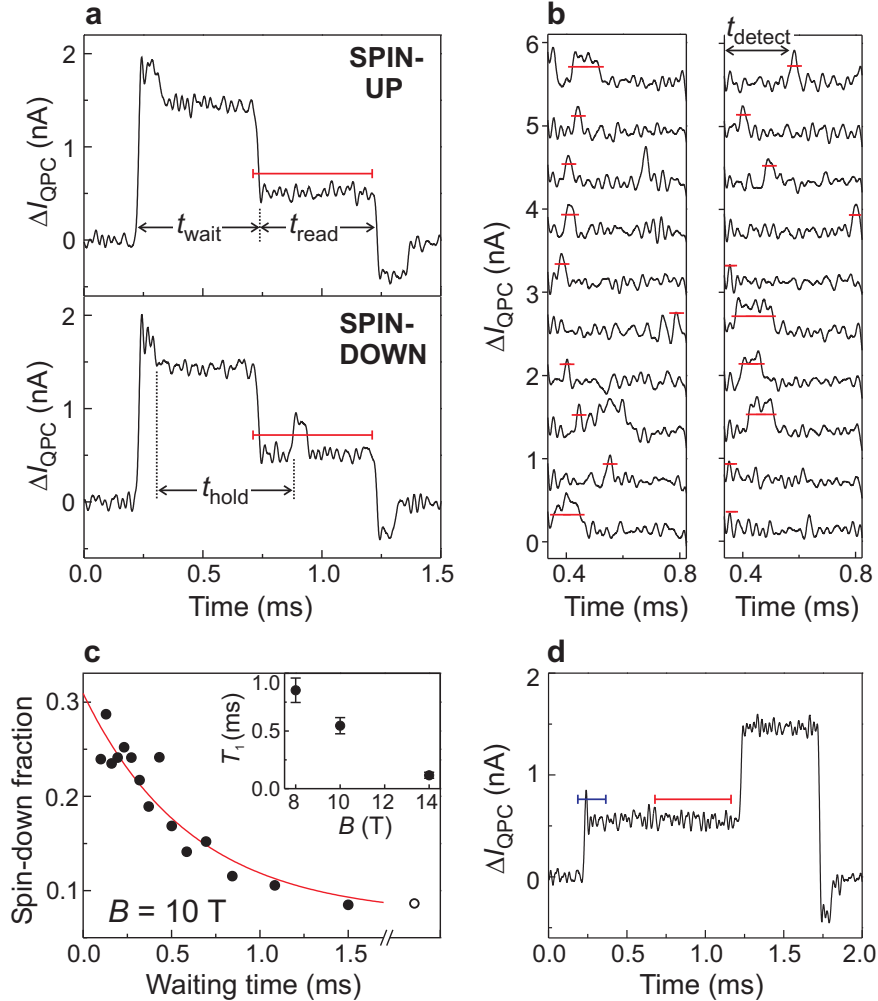}
\caption{Single-shot read-out of one electron spin. 
\textbf{(a)} Time-resolved QPC measurements. Top panel: an electron injected during $t_{wait}$ is declared `spin-up' during $t_{read}$. Bottom panel: the electron is declared `spin-down'. 
\textbf{(b)} Examples of `spin-down' traces (for $t_{wait}$ = 0.1 ms). Only the read-out segment is shown, and traces are offset for clarity. The time when $\Delta I_{QPC}$ first crosses the threshold, $t_{detect}$, is recorded to make the histogram in Fig.~\ref{fig5:fidelity}a. 
\textbf{(c)} Fraction of `spin-down' traces versus $t_{wait}$, out of 625 traces for each waiting time. Open circle: spin-down fraction using modified pulse shape (d). Red solid line: exponential fit to the data. Inset: $T_1$ versus $B$. 
\textbf{(d)} Typical QPC-signal for a `reversed' pulse, with the same amplitudes as in Fig.~\ref{fig5:diagrams}a, but the order of the two stages reversed, so that only a spin-$\ua$ electron can be injected. The fraction of traces nevertheless declared `spin-down' gives an independent measure of the `dark count' probability. This fraction is plotted as the open circle in (c) and is used in the exponential fit with an associated value of $t_{wait}$ = 10 ms (i.e. $>> T_1$). The blue threshold is used in Fig.~\ref{fig5:fidelity}b}. 
\label{fig5:singleshot}
\end{figure}

Figure~\ref{fig5:singleshot}a shows typical experimental traces of the pulse-response recorded after proper tuning of the DC gate voltages (see Fig.~\ref{fig5:tuning}). We emphasize that each trace involves injecting one particular electron on the dot and subsequently measuring its spin state. Each trace is therefore a single-shot measurement. The traces we obtain fall into two different classes; most traces qualitatively resemble the one in the top panel of Fig.~\ref{fig5:singleshot}a, some resemble the one in the bottom panel. These two typical traces indeed correspond to the signals expected for a spin-$\ua$ and a spin-$\da$ electron (Fig.~\ref{fig5:diagrams}b), a strong indication that the electron in the top panel of Fig.~\ref{fig5:singleshot}a was spin-$\ua$ and in the bottom panel spin-$\da$. The distinct signature of the two types of responses in $\Delta I_{QPC}$ permits a simple criterion for identifying the spin~\cite{note}: if $\Delta I_{QPC}$ goes above the threshold value (red line in Fig.~\ref{fig5:singleshot}a and chosen as explained below), we declare the electron `spin-down'; otherwise we declare it `spin-up'. Fig.~\ref{fig5:singleshot}b shows the read-out section of twenty more `spin-down' traces, to illustrate the stochastic nature of the tunnel events.

The random injection of spin-$\ua$ and spin-$\da$ electrons prevents us from checking the outcome of any individual measurement. Therefore, in order to further establish the correspondence between the actual spin state and the outcome of our spin measurement, we change the probability to have a spin-$\da$ at the beginning of the read-out stage, and compare this with the fraction of traces in which the electron is declared `spin-down'. As $t_{wait}$ is increased, the time between injection and read-out, $t_{hold}$, will vary accordingly ($t_{hold} \approx t_{wait}$). The probability for the spin to be $\da$ at the start of $t_{read}$ will thus decay exponentially to zero, since electrons in the excited spin state will relax to the ground state ($k_B T << \Delta E_Z$). For a set of 15 values of $t_{wait}$ we take 625 traces for each $t_{wait}$, and count the fraction of traces in which the electron is declared `spin-down' (Fig.~\ref{fig5:singleshot}c). The fact that the expected exponential decay is clearly reflected in the data confirms the validity of the spin read-out procedure.

\begin{figure}[t]
\centering
\includegraphics[width=11.5cm, clip=true]{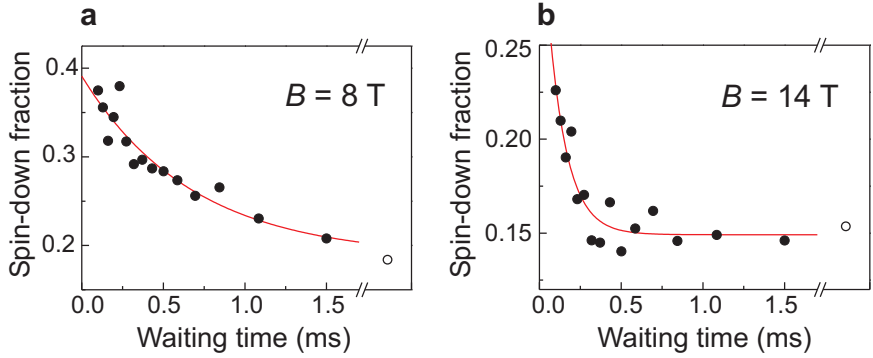}
\caption{Measurement of the spin-relaxation time as in Fig.~\ref{fig5:singleshot}c, but at different magnetic fields. Averaging the results of an exponential fit (as shown) over three similar measurements yields 
\textbf{(a)}, $T_1 = (0.85 \pm 0.11)$ ms at 8 T and 
\textbf{(b)},  $T_1 = (0.12 \pm 0.03)$ ms at 14 T.}
\label{fig5:relaxation}
\end{figure}

We extract a single-spin energy relaxation time, $T_1$, from fitting the datapoints in Fig.~\ref{fig5:singleshot}c (and two other similar measurements) to $\alpha + C \exp(-t_{wait}/T_1)$, and obtain an average value of $T_1 \approx (0.55 \pm 0.07)$ ms at 10 Tesla. This is an order of magnitude longer than the lower bound on $T_1$ established earlier \cite{rh03}, and clearly longer than the time needed for the spin measurement (of order $1/\Gamma_{\da} \approx 0.11$ ms). A similar experiment at 8 Tesla gives $T_1 \approx (0.85 \pm 0.11)$ ms and at 14 Tesla we find $T_1 \approx (0.12 \pm 0.03)$ ms (Fig.~\ref{fig5:relaxation}). More experiments are needed in order to test the theoretical prediction that relaxation at high magnetic fields is dominated by spin-orbit interactions \cite{khaetskii,golovach,woods}, with smaller contributions resulting from hyperfine interactions with the nuclear spins \cite{khaetskii,erlingsson} (cotunnelling is insignificant given the very small tunnel rates). We note that the obtained values for $T_1$ refer to our entire device under active operation: i.e. a single spin in a quantum dot subject to continuous charge detection by a QPC.


\begin{figure}[t]
\centering
\includegraphics[width=11.5cm, clip=true]{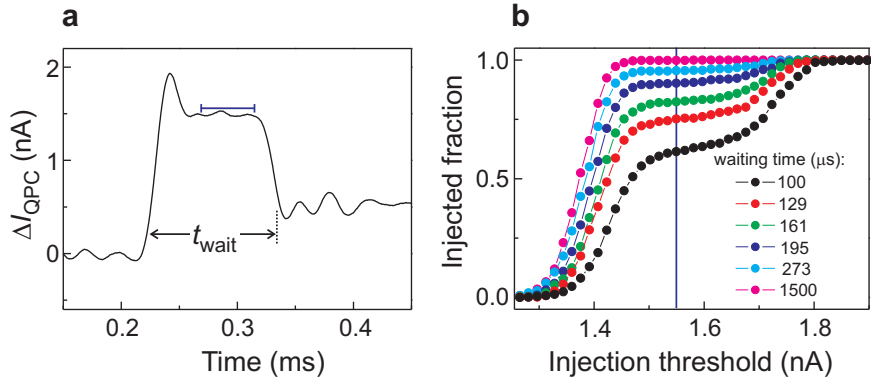}
\caption{Setting the injection threshold. 
\textbf{(a)} Example of QPC-signal for the shortest waiting time used (0.1 ms). The blue horizontal line indicates the injection threshold. Injection is declared successful if the QPC-signal is below the injection threshold for a  part or all of the last 45 $\mu$s before the end of the injection stage ($t_{wait}$). Traces in which injection was not successful, i.e. no electron was injected during $t_{wait}$, are disregarded. 
\textbf{(b)} Fraction of traces in which injection was successful, out of a total of 625 taken for each waiting time. The threshold chosen for analysing all data is indicated by the vertical blue line.}
\label{fig5:injection}
\end{figure}

\section{Measurement fidelity}
\label{fidelity}
\begin{figure}[t]
\centering
\includegraphics[width=11.5cm, clip=true]{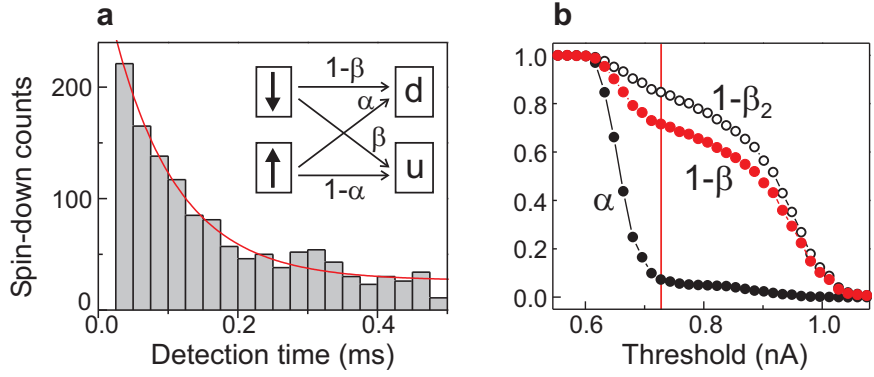}
\caption{Measurement fidelity. 
\textbf{(a)} Histogram showing the distribution of detection times, $t_{detect}$, in the read-out stage (see Fig.~\ref{fig5:singleshot}b for a definition of $t_{detect}$). The exponential decay is due to spin-$\da$ electrons tunnelling out of the dot (rate $=\Gamma_{\da}$) and due to spin flips during the read-out stage (rate $= 1/T_1$). Solid line: exponential fit with a decay time ($\Gamma_{\da} + 1/T_1)^{-1}$ of 0.09 ms. Given that $T_1$ = 0.55 ms, this yields $\Gamma_{\da}^{-1} \approx 0.11$ ms. Inset: fidelity parameters. A spin-$\ua$ electron is declared `up' or `down' with probability $1-\alpha$ or $\alpha$, respectively. A spin-$\da$ electron is declared `down' (d) or `up' (u) with probability $1-\beta$ or $\beta$, respectively. 
\textbf{(b)} Filled dark circles represent $\alpha$, obtained from the saturation value of exponential fits as in Fig.~\ref{fig5:singleshot}c for different values of the read-out threshold. A current of 0.54 nA (0.91 nA) corresponds to the average value of $\Delta I_{QPC}$ when the dot is occupied (empty) during $t_{read}$. Open circles: measured fraction of `reverse-pulse' traces in which $\Delta I_{QPC}$ crosses the injection threshold (blue line in Fig.~\ref{fig5:singleshot}d). This fraction approximates $1-\beta_2$, where $\beta_2$ is the probability of identifying a spin-$\da$ electron as `spin-up' due to the finite bandwidth of the measurement setup. Red circles: total fidelity for the spin-$\da$ state, $1-\beta$, calculated using $\beta_1 = 0.17$. The vertical red line indicates the threshold for which the visibility $1-\alpha-\beta$ (difference between filled circles and open squares) is maximal. This threshold value of 0.73 nA is used in the analysis of Fig.~\ref{fig5:singleshot}.}
\label{fig5:fidelity}
\end{figure}

For applications in quantum information processing it is important to know the accuracy, or fidelity, of the single-shot spin read-out. The measurement fidelity is characterised by two parameters, $\alpha$ and $\beta$ (inset to Fig.~\ref{fig5:fidelity}a), which we now determine for the data taken at 10 T. 

The parameter $\alpha$ corresponds to the probability that the QPC-current exceeds the threshold even though the electron was actually spin-$\ua$, for instance due to thermally activated tunnelling or electrical noise (similar to `dark counts' in a photon detector). The combined probability for such processes is given by the saturation value of the exponential fit in Fig.~\ref{fig5:singleshot}c, $\alpha$, which depends on the value of the threshold current. We analyse the data in Fig.~\ref{fig5:singleshot}c using different thresholds, and plot $\alpha$ in Fig.~\ref{fig5:fidelity}b. 

The parameter $\beta$ corresponds to the probability that the QPC-current stays below the threshold even though the electron was actually spin-$\da$ at the start of the read-out stage. Unlike $\alpha$, $\beta$ cannot be extracted directly from the exponential fit (note that the fit parameter $C = p (1-\alpha-\beta)$ contains two unknowns: $p = \Gamma_{\da}/(\Gamma_{\ua} + \Gamma_{\da})$ and $\beta$). We therefore estimate $\beta$ by analysing the two processes that contribute to it. First, a spin-$\da$ electron can relax to spin-$\ua$ before spin-to-charge conversion takes place. This occurs with probability $\beta_1 = 1/(1 + T_1 \Gamma_{\da})$. From a histogram (Fig.~\ref{fig5:fidelity}a) of the actual detection time, $t_{detect}$ (see Fig.~\ref{fig5:singleshot}b), we find $\Gamma_{\da}^{-1} \approx 0.11$ ms, yielding $\beta_1 \approx 0.17$. Second, if the spin-$\da$ electron does tunnel off the dot but is replaced by a spin-$\ua$ electron within about 8 $\mu$s, the resulting QPC-step is too small to be detected. The probability that a step is missed, $\beta_2$, depends on the value of the threshold. It can be determined by applying a modified (`reversed') pulse (Fig.~\ref{fig5:singleshot}d). For such a pulse, we know that in each trace an electron is injected in the dot, so there should always be a step at the start of the pulse. The fraction of traces in which this step is nevertheless missed, i.e. $\Delta I_{QPC}$ stays below the threshold (blue line in Fig.~\ref{fig5:singleshot}d), gives $\beta_2$. We plot $1-\beta_2$ in Fig.~\ref{fig5:fidelity}b (open circles). The resulting total fidelity for spin-$\da$ is given by $1-\beta \approx (1-\beta_1)(1-\beta_2)+(\alpha \beta_1)$. The last term accounts for the case when  a spin-$\da$ electron is flipped to spin-$\ua$, but there is nevertheless a step in $\Delta I_{QPC}$ due to the dark-count mechanism~\cite{note2}. In Fig.~\ref{fig5:fidelity}b we also plot the extracted value of $1-\beta$ as a function of the threshold. 

We now choose the optimal value of the threshold as the one for which the visibility $1- \alpha - \beta$ is maximal (red vertical line in Fig.~\ref{fig5:fidelity}b). For this setting, $\alpha \approx 0.07$, $\beta_1 \approx 0.17$, $\beta_2 \approx 0.15$, so the measurement fidelity for the spin-$\ua$ and the spin-$\da$ state is $\sim0.93$ and $\sim0.72$ respectively. The measurement visibility in a single-shot measurement is thus at present $65\%$.

Significant improvements in the spin measurement visibility can be made by lowering the electron temperature (smaller $\alpha$) and especially by making the charge measurement faster (smaller $\beta$). Already, the demonstration of single-shot spin read-out and the observation of $T_1$ of order 1 ms are encouraging results for the use of electron spins as quantum bits.\\

We thank D. P. DiVincenzo, H. A. Engel, T. Fujisawa, V. Golovach, Y. Hirayama, D. Loss, T. Saku, R. Schouten, and S. Tarucha for technical support and helpful discussions. This work was supported by a Specially Promoted Research Grant-in-Aid from the Japanese Ministry of Education, the DARPA-QUIST program, the ONR, the EU-RTN network on spintronics, and the Dutch Organisation for Fundamental Research on Matter (FOM).



\begin{thebibliography}{99}
\bibitem{note}
The automated data analysis procedure first corrects for the offset of each trace. This offset, resulting from low-frequency interference signals and charge switches, is found by making a histogram of the QPC current during the read-out stage of a particular trace. The histogram typically displays a peak due to fluctuations around the average value corresponding to an occupied dot. The center of a gaussian fit to the histogram gives the offset. Then each trace is checked to make sure that an electron was injected during the injection stage, by evaluating if the signal goes below the injection threshold (blue horizontal line in Fig.~\ref{fig5:injection}a). If not, the trace is disregarded. Finally, to determine if a trace corresponds to `spin-up' or `spin-down', we disregard all points that lie below the previous point (since these could correspond to points on the falling pulse flank at the end of the injection stage), and check if any of the remaining points are above the threshold.
\bibitem{note2}
Let us assume there is a spin-$\da$ electron on the dot at the start of the read-out stage. The probability that the $\da$-electron tunnels out (i.e. that it does not relax to spin-$\ua$) is given by $1-\beta_1$. The probability that this tunnel event is detected (i.e. is not too fast) is given by $1-\beta_2$. Therefore, the probability that a spin-$\da$ electron tunnels out and is detected, is $(1-\beta_1)(1-\beta_2)$. In addition, there is the possibility that the $\da$-electron relaxes, with probability $\beta_1$, but a step in the QPC signal is nevertheless detected, with probability $\alpha$, due to the `dark count' mechanism. Therefore, the total probability that a spin-$\da$ electron is declared `spin-down' is given by $(1-\beta_1)(1-\beta_2) + (\alpha \beta_1)$ approximately.
\end{thebibliography}
\end{document}